\documentclass[rnote]{aa}  

\usepackage{graphicx}
\usepackage{txfonts}
\newcommand{\HI}{H{\,\small I}}

\newcommand{\ltsima} {$\; \buildrel < \over \sim \;$}
\newcommand{\gtsima} {$\; \buildrel > \over \sim \;$}
\newcommand{\lta} {\lower.5ex\hbox{\ltsima}}
\newcommand{\gta} {\lower.5ex\hbox{\gtsima}}

\newcommand{\kms}{km\ s$^{-1}$}

\begin{document}

   \title{From galaxy-scale fueling to nuclear-scale feedback}
   \subtitle{The merger-state of radio galaxies 3C\,293, 3C\,305 $\&$ 4C\,12.50}

   \author{B.H.C. Emonts
          \inst{1}\thanks{Marie Curie Fellow}
          \and R. Morganti
          \inst{2,3}
          \and M. Villar-Mart\'{i}n
          \inst{1,4}
          \and J. Hodgson
          \inst{5}
          \and E. Brogt
          \inst{6}
          \and C.N. Tadhunter
          \inst{7}
          \and E. Mahony
          \inst{8,9}
          \and T.A. Oosterloo
          \inst{2,3}
          }

   \institute{Centro de Astrobiolog\'{i}a (INTA-CSIC), Ctra de Torrej\'{o}n a Ajalvir, km 4, 28850 Torrej\'{o}n de Ardoz, Madrid Spain\\
              \email{bjornemonts@gmail.com}
         \and
         ASTRON, the Netherlands Institute for Radio Astronomy, Postbus 2, 7990 AA, Dwingeloo, The Netherlands
         \and
         Kapteyn Astronomical Institute, University of Groningen, P.O. Box 800, 9700 AV Groningen, The Netherlands
         \and
         Astro-UAM, UAM, Unidad Asociada CSIC, Facultad de Ciencias, Campus de Cantoblanco, E-28049, Madrid, Spain
         \and
         Korea Astronomy and Space Science Institute, 776 Daedeokdae-ro, Yuseong-gu, Daejeon 34055, Korea
         \and
         Academic Services Group, University of Canterbury, Christchurch, Private Bag 4800, Christchurch 8140, New Zealand
         \and
         Department of Physics and Astronomy, University of Sheffield, Sheffield S3 7RH, UK
         \and
         Sydney Institute for Astronomy, School of Physics A28, University of Sydney, NSW 2006, Australia
         \and
         ARC Centre of Excellence for All-Sky Astrophysics (CAASTRO)
             }

   \date{}

 
  \abstract
{Powerful radio galaxies are often associated with gas-rich galaxy mergers. These mergers may provide the fuel to trigger starburst and active galactic nuclear (AGN) activity. In this Research Note, we study the host galaxies of three seemingly young or re-started radio sources that drive fast outflows of cool neutral hydrogen (\HI) gas, namely 3C\,293, 3C\,305 and 4C\,12.50 (PKS\,1345+12). Our aim is to link the feedback processes in the central kpc-scale region with new information on the distribution of stars and gas at scales of the galaxy. For this, we use deep optical V-band imaging of the host galaxies, complemented with \HI\ emission-line observations to study their gaseous environments. We find prominent optical tidal features in all three radio galaxies, which confirm previous claims that 3C\,293, 3C\,305 and 4C\,12.50 have been involved in a recent galaxy merger or interaction. Our data show the complex morphology of the host galaxies and identify the companion galaxies that are likely involved in the merger or interaction. The radio sources appear to be (re-)triggered at a different stage of the merger; 4C\,12.50 is a pre-coalescent and possibly multiple merger, 3C\,293 is a post-coalescent merger that is undergoing a minor interaction with a close satellite galaxy, while 3C\,305 appears to be shaped by an interaction with a gas-rich companion. For 3C\,293 and 3C\,305, we do not detect \HI\ beyond the inner $\sim$30-45 kpc region, which shows that the bulk of the cold gas is concentrated within the host galaxy, rather than along the widespread tidal features.}


   \maketitle
%

\section{Introduction}
\label{sec:intro}

Deep optical broadband imaging has revealed that powerful radio galaxies are often associated with gas-rich galaxy mergers or interactions \citep[][see also work by \citealt{hec86}, \citealt{smi89}, \citealt{roc00}, \citealt{sab13}]{ram12}. Furthermore, powerful radio galaxies often show young stellar populations and contain dust masses that link them to merger activity \citep{tad11,tad14a}. These gas-rich mergers and interactions are believed to deposit the cold material that is needed to fuel both starburst activity \citep[e.g.,][]{dim07} and the powerful radio sources \citep[e.g.,][]{har07}. 

Although the majority of early-type galaxies outside clusters contain at least modest amounts of cold gas \citep[few $\times$ 10$^{6-9}$ M$_{\odot}$;][]{mor06,oos10,ser12}, the host galaxies of small radio sources can be particularly rich in \HI\ \citep{emo07}. Moreover, young or recently re-started radio sources often contain large amounts of \HI\ gas in their central regions \citep{pil03,ver03,gup06a,gup06,cha11,cha13,ger14}. On average, these central regions appear much richer in \HI\ than those occupied by extended radio sources \citep{gup06,cha13,ger14}. Moreover, the \HI\ profiles associated with young radio sources more often show blue wings, which suggests that these sources drive gas outflows as they clear their way through the rich ambient medium of the host galaxy \citep{cha11,ger15}. 

Particularly interesting are young or re-started radio sources that have been observed to drive very fast ($\sim$1000 \kms) outflows of cool neutral and cold molecular gas from the central kpc-scale region \citep{oos00,mor03,mor05,mor05sample,das12,mor13sci,mor13,mah13}. The energy released by these jet-driven outflows approaches values needed to clear the central region of gas and quench star formation \citep{mor13,mah15}, certainly when taking into account that part of the kinetic energy of the jets is put into heating of the cold gas \citep{gui12,har12,tad14,das14,lan15}. Thus, the jet-induced feedback may influence the evolution of these galaxies.

In this paper, we further investigate the link between the nuclear feedback phenomena and large-scale merger processes for three of these seemingly young or re-started radio galaxies with fast outflows of cold gas, namely 3C\,293, 3C\,305 and 4C\,12.50 (PKS\,1345+12). 
The cold outflows in these systems were identified through blueshifted \HI\ seen in absorption against the strong radio continuum \citep{mor03,mor05,mor13sci,mah13}. Previous optical imaging revealed evidence for tidal debris from a galaxy merger or interaction in these systems \citep{hec86}. We here present more detailed optical imaging, combined with optical spectroscopy and deep \HI\ observations. Our aim is to identify the stage of merger or interaction in these systems. With this, we complement the studies of the feedback processes in the kpc-scale central region with new information on the distribution of stars and gas at scales of the host-galaxy environment. 


Throughout this paper we will assume $H_{0} = 71$\,\kms\,Mpc$^{-1}$, $\Omega_{\rm M} = 0.27$ and $\Omega_{\Lambda} = 0.73$.

\section{Observations}
\label{sec:observations}

\subsection{Optical imaging}
\label{sec:obsoptical}

Deep optical V-band imaging of the three radio galaxies was done in March 2007 and April 2008 and at the Hiltner 2.4m telescope of the Michigan-Dartmouth-MIT (MDM) observatory, Kitt Peak, Arizona (USA). Imaging was done using the Echelle CCD, with a field-of-view (F.o.V.) of $9.5^{\prime} \times 9.5^{\prime}$. Observations were taken with a seeing of 1-2$^{\prime\prime}$. No absolute flux calibration was performed. Table 1 summarizes the observations.

The Image Reduction and Analysis Facility \citep[{\sc iraf};][]{tod93} was used for a standard data reduction (bias subtraction, flat-fielding, frame alignment and cosmic-ray removal). A background gradient, which was most likely introduced by minor shutter issues, had to be subtracted from the final images \citep[see][]{emo10}. Using the Karma software \citep{goo96}, a world coordinate system (accurate to within 1 arcsec) was copied from a Sloan Digital Sky Survey (SDSS) image of the same region.

An optical spectrum with a single 300 sec exposure was obtained for the object seen in the vicinity of 3C\,293 (Sect.\,\ref{sec:3C293}) using the Intermediate dispersion Spectrograph and Imaging System (ISIS) with R300B/R600R gratings on the William Herschel Telescope (WHT) on 29 June 2006. The sole purpose of this spectrum was to identify the object as a companion galaxy and to determine its redshift, hence a standard wavelength calibration was applied, but no absolute flux scale.

\begin{table}
\label{tab:table}
\caption{Optical V-band imaging.}
\centering
\begin{tabular}{lcccc}
Source & $z$ & Date & t$_{\rm int}$ (sec) & airmass \\
\hline
3C\,293        & 0.045 & 13/03/07 & 3600 ($12 \times 300$) & 1.0 \\
3C\,305        & 0.041 & 05/04/08 &  5400 ($6 \times 900$) & 1.2 \\
4C\,12.50   & 0.123 & 06/04/08 &  5400 ($6 \times 900$) & 1.1 \\
\end{tabular}
\tablefoot{t$_{\rm int}$ is the on-source integration time ($\#$ frames $\times$ time per frame).}
\end{table}

\subsection{Radio H\,I spectroscopy}
\label{sec:obsradio}

Deep 21cm radio observations were taken with the Westerbork Synthesis Radio Telescope (WSRT) to search for \HI\ emission in 3C\,293 and 3C\,305. At $z$\,=\,0.12, 4C\,12.50 is too distant to image \HI\ in emission with the WSRT. 3C\,293 was observed on 14, 15, 16, 19 May 2007 for a total of 4\,$\times$\,12h, 3C\,305 on 29 June $\&$ 3 July 2010 for 2\,$\times$\,12h. We used the 20 MHz band, with for 3C\,293 a double IF with 512 channels ($\Delta$v $\sim$ 8 \kms) and for 3C\,305 a single IF with 1024 channels ($\Delta$v $\sim$ 4 \kms). A primary calibrator (3C\,286 or CTD\,93) was used to calibrate the bandpass every hour to reach the required 1:10,000 spectral dynamic range for detecting weak \HI\ emission in the vicinity of the strong radio continuum cores (3.7 for 3C\,293 and 3.0 Jy for 3C\,305). The flux scale was set using 3C\,286. 

The data reduction was done using {\sc miriad} \citep{sau95}. After flagging, a continuum image was made by first fitting a straight line to the signal in the channels that do not contain any \HI\ line, and subsequently Fourier transforming, self-calibrating and cleaning these continuum data. A continuum image from earlier WSRT data of 3C\,293 was presented in \citet[][]{emo05}. The continuum source of 3C\,305 is unresolved. The calibration solutions of the continuum data were then transferred to the original `line+continuum' data, while the continuum model was subtracted from this data set in the uv-plane. Any residual continuum was subtracted from the line data by fitting a straight line to the line-free channels. For 3C\,293, any remaining phase and amplitude errors were corrected by a selfcalibration on the strong ($\sim$200 mJy) absorption line in the corresponding channels. We then Fourier transformed the line-data using robust +0.5 weighting \citep{bri95} and cleaned it using a mask (created by smoothing the data) to isolate the regions in which \HI\ emission or absorption was found. 

The resulting data of 3C\,293 and 3C\,305 have a beam-size of 34.9''\,$\times$\,20.5'' (PA -1.0$^{\circ}$) and 20.9''\,$\times$\, 19.8'' (PA 1.1$^{\circ}$), respectively. After binning and Hanning smoothing the data, we reached a thermal-noise limited sensitivity of 0.32 mJy\,beam$^{-1}$ per 8.2 \kms\ channel for 3C\,293 and 0.36 mJy\,beam$^{-1}$ per re-binned 17.8 \kms\ channel for 3C\,305.

\section{Results}

Figures \ref{fig:3C293}, \ref{fig:3C305} and \ref{fig:PKS1345} show the V-band imaging of 3C\,293, 3C\,305 and 4C\,12.50. Various features are presented with different scaling. For 3C\,293 and 3C\,305, the \HI\ results are also included. 

\subsection{3C\,293}
\label{sec:3C293}

   \begin{figure*}
   \centering
   \includegraphics[width=0.95\textwidth]{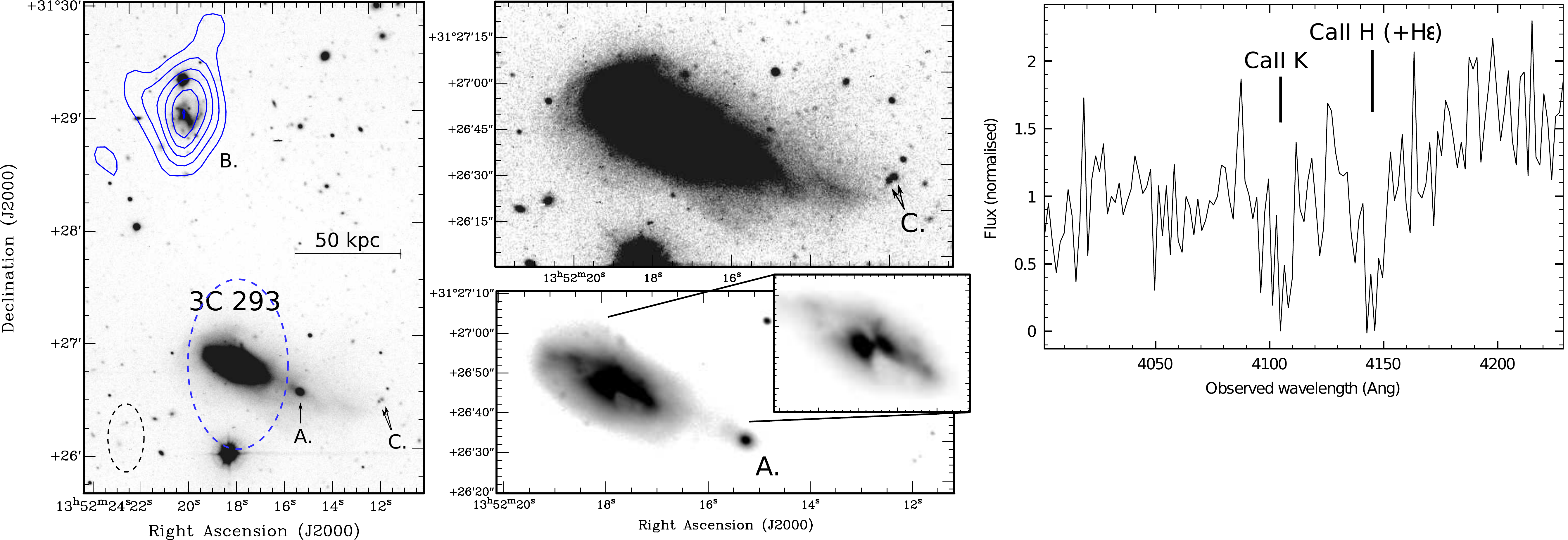}
   \caption{Optical V-band imaging of 3C\,293. {\sl Left+middle:} The various plots show the same image of 3C\,293, but with different intensity-scaling and zooming. On the left is shown the larger environment of 3C\,293, with overlaid in blue the contours of HI 21cm emission (levels: 0.04, 0.07, 0.10, 0.13, 0.16, 0.19 Jy\,bm$^{-1}$\,$\times$\,km\,s$^{-1}$). The dashed blue ellipse represents the region where the \HI\ absorption dominates \citep[see][]{mor03}. Even though this \HI\ absorption is spatially unresolved, it is so strong that it dominates over any potential \HI\ emission far beyond the FWHM of the synthesized beam (dashed black ellipse on the bottom-left). The companion galaxies A and B, and feature C, are described in the text. {\sl Right:} Optical spectrum of source A. The Ca\,II K and H absorption lines confirm that this source is a companion of 3C\,293.}
              \label{fig:3C293}
    \end{figure*}


Figure \ref{fig:3C293} shows the V-band image of 3C\,293. The high-surface-brightness inner region displays a disc-like morphology, but with prominent dust-lanes \citep[better visualized in the HST imaging by][]{bes02}. The low-surface-brightness envelope shows a distorted morphology, and a faint bridge connecting 3C\,293 with the nearby small companion `A' at 13 kpc west-south-west of the nucleus of 3C\,293. This small companion shows an elliptical, nearly unresolved, morphology. To confirm this object as a true companion, we obtained an optical spectrum (Fig.\,\ref{fig:3C293}, right). By fitting a Gaussian profile to the Ca\,II K and H features, we find that companion A has a velocity of 13171\,$\pm$\,260. This corresponds to v\,=\,-279\,$\pm$\,260 \kms\ with respect to the systemic velocity of 3C\,293, which is v$_{\rm sys}$\,=\,13,450 \kms\ \citep{emo05}. 

The low-surface-brightness envelope of the optical emission around 3C\,293 is extended in the direction of this companion A. About 70 kpc further to the west (beyond companion A) the low-surface-brightness tail shows a `zig-zag' pattern and ends in two knots (indicated as `C' in the top-right plot of Fig.\,\ref{fig:3C293}). It is interesting to speculate that these two knots could represent star forming regions at the tip of the low-surface-brightness tail.

No \HI\ emission is detected, down to a 5$\sigma$ limit of $7.0 \times 10^{19}$ cm$^{-2}$ across 100 \kms, outside the central $\sim$45 kpc region where the \HI\ absorption-signal dominates. Because of the strong \HI\ absorption, any superposed weak \HI\ emission cannot be distinguished within the central region. Part of the \HI\ absorption traces a central disk \citep{baa81,bes02,bes04}, which was imaged in CO \citep[][]{eva99,lab14} and ionized gas \citep{emo05,mah15}. A fast outflow of both \HI\ and ionized gas was reported by \citet{mor03} and \citet{emo05}, and studied in detail against the inner kpc-scale radio source by \citet{mah13,mah15}. 

Another companion `B' is found at 120 kpc distance north of 3C\,293 and detected in \HI\ emission \citep[with M$_{\rm HI}$\,=\,1.6\,$\times$\,10$^{9}$\,M$_{\odot}$ and v\,=\,13,416 \kms;][]{emo06thesis}. This companion appears to contain a face-on starforming disc with a prominent bar. There is no indication from the optical or \HI\ morphology that this companion is interacting with 3C\,293.

\subsection{3C\,305}
\label{sec:3C305}

\begin{figure*}
   \centering
   \includegraphics[width=0.95\textwidth]{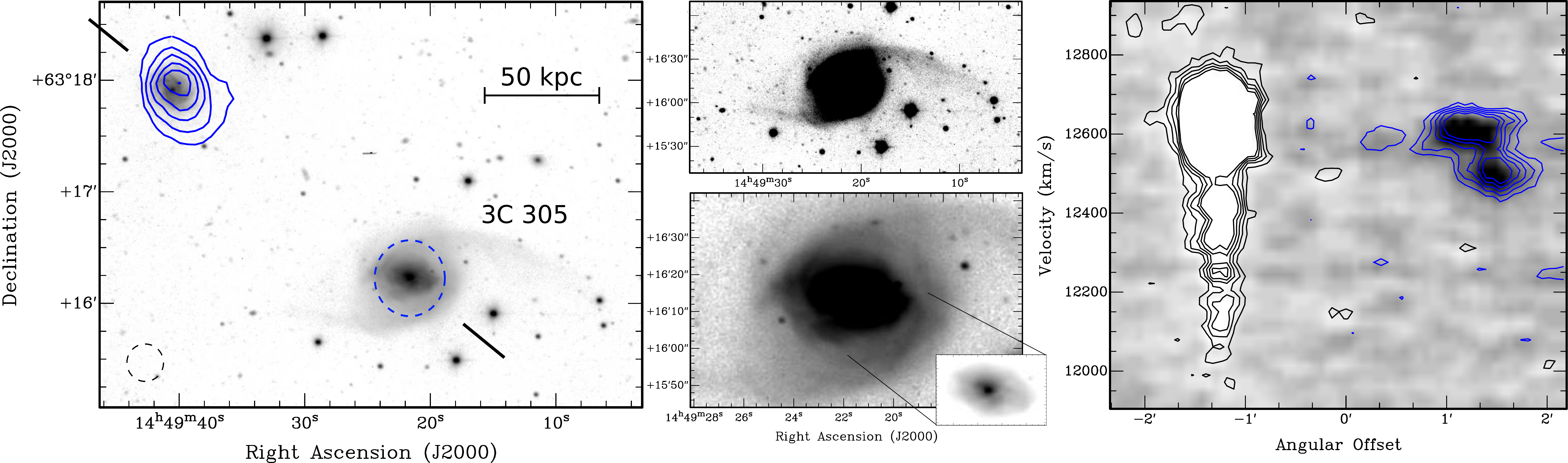}
   \caption{Optical V-band imaging of 3C\,305. {\sl Left+middle:} The various plots show the same image of 3C\,305, but with different intensity-scaling and zooming. On the left is shown the larger environment of 3C\,305, with overlaid in blue the contours of HI 21cm emission. The dashed blue circle shows the region where the \HI\ absorption dominates \citep[see][]{mor05}. Even though this \HI\ absorption is spatially unresolved, it is strong enough to dominate over any potential \HI\ emission beyond the FWHM of the synthesized beam (dashed black circle on the bottom-left). {\sl Right:} Position-velocity plot of the \HI\ emission (blue contours) and absorption (black contours) along the axis shown in the left plot. Contour levels are at -3, -2, 2, 3, 4, 5, 6 $\times$ $\sigma$, with $\sigma$\,=\,0.36 mJy beam$^{-1}$.}
              \label{fig:3C305}
    \end{figure*}

Figure \ref{fig:3C305} shows the V-band image of 3C\,305. The low-surface-brightness image shows two tidal arms stretching in east-west direction. The major axis of the galaxy seems to turn from $\sim$145$^{\circ}$ in the low-surface-brightness plot (top-middle) to $\sim$70$^{\circ}$ in the high-surface-brightness plot (bottom-middle). The inner region also shows a distorted morphology, with arc- or spiral-like isophotes. 3C\,305 hosts a single nucleus \citep{jac03}.

No \HI\ is detected in emission, down to a 5$\sigma$ limit of $2.1 \times 10^{20}$ cm$^{-2}$ across 100 \kms, outside the central $\sim$30 kpc region where the \HI\ absorption-signal dominates. As for 3C\,293, also in 3C\,305 part of this \HI\ absorption likely traces an inner gas disc \citep{jac03}, while another part is a fast gas-outflow detected against the inner radio-jet at kpc-distance from the nucleus \citep{mor05}.

North-east of 3C\,305, at 133 kpc distance, a companion galaxy is detected in \HI\ emission (M$_{\rm HI}$\,=\,4.2\,$\times$\,10$^{9}$\,M$_{\odot}$). The \HI\ velocity of this companion is v\,=\,12,560 \kms. This is very close to v$_{\rm sys}$\,=\,12,550 \kms\ for 3C\,305 \citep{mor05}. Part of the \HI\ appears marginally elongated in the direction of 3C\,305, as shown by the lowest contours in the position-velocity plot of the \HI\ gas in Fig.\,\ref{fig:3C305} (right). If confirmed, it would suggest that an interaction occurred between 3C\,305 and this gas-rich companion. It is interesting to note that the axis of the kpc-scale radio source in 3C\,305 is aligned not only parallel to the nuclear dust lane \citep{jac03}, but also in the direction of this gas-rich companion. This is reminiscent of similar alignments seen in the nearby radio galaxies NGC\,612 and B2\,0722+30 \citep[][]{emo09}. However, we cannot rule out that this is a chance-alignment and the radio jet in 3C\,305 is merely perpendicular to the large-scale disc (as seen in other radio galaxies by \citealt{kav15}).

\subsection{4C\,12.50 (PKS\,1345+12)}
\label{sec:PKS1345}

\begin{figure}
   \centering
   \includegraphics[width=0.47\textwidth]{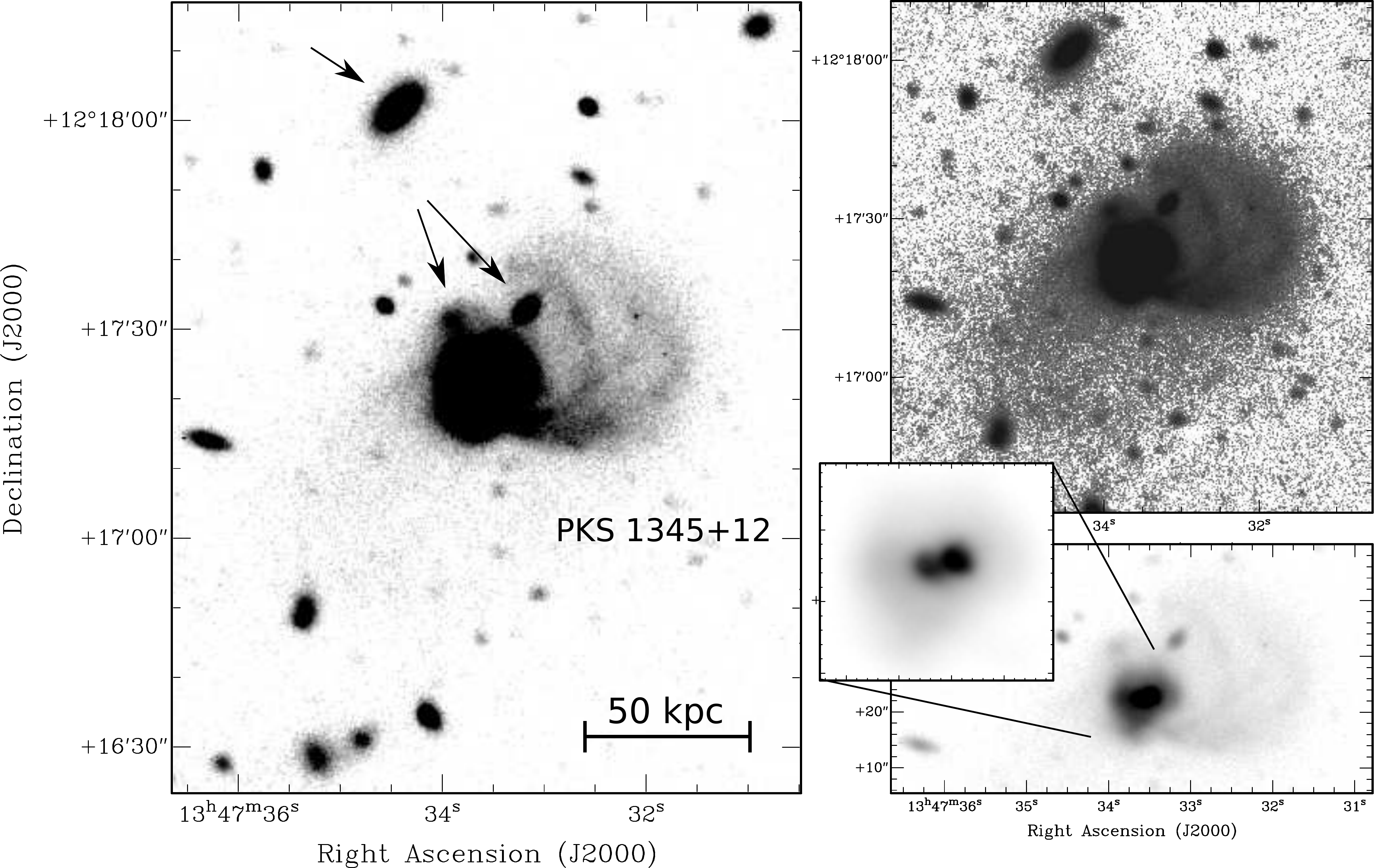}
   \caption{Optical V-band imaging of 4C\,12.50 (PKS\,1345+12). The various plots show the same image of 4C\,12.50, but with different intensity-scaling and zooming. Potential companion galaxies discussed in the text are marked with arrows in the left plot.}
              \label{fig:PKS1345}
    \end{figure}

Figure \ref{fig:PKS1345} shows the V-band image of 4C\,12.50 (PKS\,1345+12). It reveals an apparent multiple merger system. The core of 4C\,12.50 consists of a double nucleus, with a separation of 1.6$''$ or 3.5 kpc (as previously found by \citealt{hec86} and \citealt{eva99}). At the south-western end of the host galaxy, a double tidal-tail stretches $\sim$60 kpc north, encircling roughly 30$\%$ of the host galaxy. This double tidal-tail stretches in the direction of an apparent bright companion $\sim$92 kpc north of 4C\,12.50. On the opposite side of 4C\,12.50 there is a broad fan of emission that stretches $\sim$85 kpc towards the south-east. 4C\,12.50 resides in a dense environment. Two apparent close companion galaxies lie $\sim$20 kpc north of the host galaxy and may be involved in the merger (see arrows in Fig.\,\ref{fig:PKS1345}). However, spectroscopic redshifts are required to confirm their association with 4C\,12.50.

At $z$\,=\,0.12, 4C\,12.50 is too distant to image \HI\ in emission with the WSRT, but the central \HI\ absorption has been studied in great detail \citep{mir89,mor04,mor13sci}. Against the tip of the southern pc-scale radio jet, VLBI observations revealed a very broad \HI\ absorption feature, which is blueshifted by $\sim$1000 \kms \citep{mor13sci}. This broad feature has also been seen in CO by \citet{das12}. In addition, a deep \HI\ absorption component occurs against the northern VLBI jet, and could represent a cloud within the central disc that was detected in CO by \citet{eva99}. Diffuse, faint radio emission is found at larger scales, which is likely a relic source from a past episode of activity \citep{sta05}.

\section{Discussion and conclusions}

The deep optical imaging that we presented of 3C\,293, 3C\,305 4C\,12.50 reveals galaxy-morphologies that agree with those previously shown by \citealt{hec86} (also \citealt{hec82}, \citealt{bre84}). However, our data show in more detail a multitude of subtle features. Specifically, combined with our optical spectroscopy and \HI\ imaging, we obtain the following new insights on these three radio galaxies:\\
\vspace{-2mm}\\
$\bullet$ 3C\,293: We confirm the presence and redshift of a close companion at $\sim$13 kpc distance, beyond which a stellar tail with a 'zig-zag' morphology ends in two possible starforming regions.\\ 
\vspace{-2mm}\\
$\bullet$ 3C\,305: The major axis of the host galaxy seems to turn by $\sim$75$^{\circ}$ from the faint outer envelope with the two tidal arms to the bright inner region. The inner region shows arc- or spiral-like isophotes. There is a tentative indication that a faint \HI\ feature stretches from an \HI-rich companion towards 3C\,305, along the direction of the radio axis.\\
\vspace{-2mm}\\
$\bullet$ 4C\,12.50: A double tidal-tail stretches in the direction of a bright companion galaxy. On the opposite side, a faint but broad fan of emission stretches up to $\sim$85 kpc from the galaxy.\\
\vspace{-2mm}\\
Our results also reveal that, for 3C\,293 and 3C\,305, no \HI\ has been detected along the stellar tidal debris beyond the central region where the \HI\ absorption-signal dominates. However, we note that sensitive \HI\ observations with a spatial resolution that better matches the extended optical features are needed to better study any potential widespread \HI\ gas in emission.

\subsection{Merger state and AGN triggering}

Our results indicate that all three systems have been involved in a recent galaxy merger or interaction, which is consistent with the presence of young or post-starburst stellar populations across these systems \citep{tad05}. 3C\,305 was classified by \citet{hec86} as a post-merger system. However, contrary to the claim by \citet{hec86} that there are no close companion galaxies that could have provoked the observed morphological disturbances, we show that 3C\,305 appears to be involved in a galaxy-encounter with a gas-rich companion. This resembles the case of NGC\,6872, where two tidal arms were formed by the low-inclination, prograde passage of companion IC\,4970 \citep[][see also \citealt{mih93}]{hor07}. If a similar passage created the tidal arms in 3C\,305, and assuming that the companion has an average velocity of $\sim$200 \kms\ with respect to 3C\,305, then the time since closest approach must have occurred $\sim$650\,Myr ago. This time-scale is similar to the 0.5\,$-$\,1 Gyr age of the young stellar population across 3C\,305 \citep{tad05}. It could suggest that a galaxy-wide starburst was triggered at the time of closest approach between the two systems, which is consistent with simulations \citep[e.g.,][]{pei10}. Additional observations are needed to investigate this. For 3C\,293, the distorted optical morphology and single nucleus \citep{flo06} suggest that this is a post-merger system which is currently experiencing a minor interaction with the close companion 13\,kpc towards the west-south-west. 4C\,12.50 shows clear evidence from its double nucleus and multiple tidal features that it is currently involved in an ongoing major --possibly multiple-- merger event. This is consistent with its high IR luminosity ($L_{\rm IR} \sim 2 \times 10^{12} L_{\odot}$, that is, in the regime of ultra-luminous infrared galaxies; \citealt{san88}). These results indicate that 3C\,305, 3C\,293 and 4C\,12.50 are likely at different stages in the merger process.

Interestingly, this also implies that the radio sources in 4C\,12.50, 3C\,305 and 3C\,293 could have been triggered at different stages during the merger. After all, 4C\,12.50, 3C\,305 and 3C\,293 are classified as a Gigahertz Peaked Spectrum (GPS), Compact Steep Spectrum (CSS) and Steep Spectrum Core (SSC) source, respectively, which likely represent young or re-started radio sources \citep[$\le$10$^{6}$ yr;][]{fan95,ode98,shu12,col16}. This different stage of triggering would be consistent with optical and IR studies, which indicate that a radio source can in general occur at any time during the merger process, either before or after the individual nuclei coalesce \citep{ram11,tad11,dic12}. We caution that, while both 3C\,293 and 4C\,12.50 have radio cores with an estimated age of $\sim$3\,$\times$\,10$^{4}$ yr \citep{aku96,ode00}, they also contain extended radio lobes \citep[e.g.,][]{bes04,sta05}. Although we assume that these outer lobes represent a previous episode of activity, we cannot rule out that the radio cores are older than they appear from their spectral-index measurements \citep[see][]{blu00}, for example, because they have been frustrated by a dense inter-stellar medium \citep[e.g.,][]{ode91}. If the radio cores are actively feeding the much larger and older radio structures, then the moment of triggering is uncertain. However, we find this scenario less likely, because 3C\,293, 3C\,305 and 4C\,12.50 are among the most powerful steep-spectrum sources at low-$z$ and display among the most extreme jet-ISM interactions known in the low-$z$ Universe. Such features are often associated with young or re-started radio sources that are trying to plough their way through the ambient ISM \citep[Sect.\,\ref{sec:intro}; see also][]{hol08,hol11}.

Concluding, our results support the notion that galaxy mergers and interactions are the likely mechanisms for accumulating cold gas in the central few kpc of these systems, and possibly also for (re-)triggering the powerful radio sources that violently interact with this dense ISM.

\begin{acknowledgements}
We thank Jacqueline van Gorkom for her help and useful discussions. We also thank the referee, Stanislav Shabala, for useful feedback that improved this paper. BE thanks Columbia University for its hospitality during part of this project. The research leading to these results has received funding from the People Programme (Marie Curie Actions) of the European Union's Seventh Framework Programme FP7/2007-2013/ under REA grant agreement n$^{\circ}$\,624351. RM gratefully acknowledges support from the European Research Council under the European Union's Seventh Framework Programme (FP/2007-2013) /ERC Advanced Grant RADIOLIFE-320745. MVM acknowledges support from the Spanish Ministerio de Econom\'{i}a y Competitividad through the grant AYA2012- 32295. EKM acknowledges support from the Australian Research Council Centre of Excellence for All-sky Astrophysics (CAASTRO), through project number CE110001020. Data were obtained using the 2.4m Hiltner Telescope of the Michigan-Dartmouth-MIT (MDM) Observatory, owned and operated by a consortium of the University of Michigan, Dartmouth College, Ohio State University, Columbia University and Ohio University. The Westerbork Synthesis Radio Telescope is operated by ASTRON (Netherlands Institute for Radio Astronomy) with support from the Netherlands Foundation for Scientific Research (NWO). The Wiliam Herschel Telescope is operated on the island of La Palma by the Isaac Newton Group in the Spanish Observatorio del Roque de los Muchachos of the Instituto de Astrofísica de Canarias.
\end{acknowledgements}


\end{document}